\documentclass[10pt,draft]{dis03}
\usepackage{epsf,amsmath}

\newcommand{\nn}{\nonumber}
\newcommand{\be}{\begin{equation}}
\newcommand{\ee}{\end{equation}}
\newcommand{\ba}{\begin{eqnarray}}
\newcommand{\ea}{\end{eqnarray}}

\def\als{\alpha_s}

\def\gev{\,{\rm GeV}}
\def\vk{{\bf k}_{\perp}}

\def\vbs{{\bf b}}
\newcommand{\LQCD}{\Lambda_{\rm{QCD}}}
\def\xb{\overline{x}}
\def\xbj{x_{\rm Bj}}

\begin{document}

\title{Leptoproduction of  Vector Mesons at Small $\xbj$
and Generalized Parton Distributions
\thanks{This work is
supported  in part by the Russian Foundation for Basic Research,
Grant 03-02-16816,  by the Deutsche Forschungsgemeinschaft and by
the Heisenberg-Landau program}}

\author{S.V.\ Goloskokov\\
BLTP, Joint Institute for Nuclear Research \\
Dubna 141980, Moscow region, Russia\\
E-mail: goloskkv@thsun1.jinr.ru\\
P.Kroll\\
Fachbereich Physik, Universit\"at Wuppertal, \\
D-42097 Wuppertal, Germany\\
E-mail: kroll@physik.uni-wuppertal.de\\
B.Postler\\
Stanford Linear Accelerator Center\\
2575 Sand Hill Road, Menlo Park, CA 94025, USA\\
E-mail: postler@SLAC.Stanford.EDU} \maketitle

\begin{abstract}
\noindent Deep virtual vector meson leptoproduction at small
$\xbj$ is analyzed in the framework of generalized parton
distributions. It is shown that the inclusion of transverse
degrees of freedom leads to reasonable agreement with the cross
section data for light  mesons.
\end{abstract}

In this article we report on an investigation of vector meson
leptoproduction off protons at small $\xbj$ and large photon
virtuality, $Q^2$. In this kinematical region the process
factorizes \cite{col96} into a hard subprocess, meson
leptoproduction off partons, and a soft proton matrix element
which represents a gluonic generalized parton distribution (GPD)
\cite{rad96}. At large $Q^2$, the cross section is dominated by
photon-meson transitions where both the particles are
longitudinally polarized. Other transitions are suppressed by at
least $1/Q$ as compared to the leading-twist one, $\gamma_L^*\to
V_L$. It has, however, turned out that the leading-twist
contribution to the cross section is substantially exceeds
experiment \cite{mpw}. This parallels observations made within the
two-gluon exchange model. As argued in \cite{fra95} transverse
momentum effects in the photon wave function provides sufficient
suppression of the two-gluon exchange amplitude in order to
achieve agreement with experiment. We remark that GPD corrections
to the two-gluon exchange model have been estimated in \cite{rys}.

Motivated by experience with meson form factors \cite{li92,jak93} and
with meson leptoproduction at large $\xbj$ \cite{vander} we are going to modify
the leading-twist GPD approach by the inclusion of transverse degrees
of freedom and Sudakov suppressions in the subprocess in order to
suppress the contributions from the end-point regions where one
of the partons entering the meson wave function becomes soft and,
hence, factorization breaks down.

As shown in \cite{col96}, to leading-twist accuracy, the
$\gamma_L\, p \to V_L \, p$ amplitude reads
\be
{\cal M}^{V(g)}_{0+,0 +} = e \, {\cal C}_V\, \sqrt{1-\xi^2}\,(1+\xi)\,
          \int_0^1 d\xb\; \frac{{\cal H}^{(g)}_{0+,0 +}\; H^g(\xb,\xi,t)}
        {(\xb+\xi) (\xb-\xi + i\hat{\varepsilon})}\,,
\label{amp-nf}
\ee
at small $\xbj$. The skewness $\xi$ is approximately given by
$\xi \simeq \xbj/2 \simeq Q^2/(2W^2)$ in the kinematical region of
interested here ($W$ is c.m. energy of the $Vp$ system). Examples of
flavor factors are
\be
 {\cal C}_{\rho}=1/\sqrt{2}\,, \qquad  {\cal C}_{\phi}=-1/3\,.
\label{flavor}
\ee
To leading order
(one-gluon exchange) the amplitude for the partonic subprocess,
$\gamma_L g\to V_L g$, acquires the form
 \be
{\cal H}^{(g)}_{0+,0 +}\, = \,\frac{4\pi \als(Q/2) f_V}{N_c Q}
           \,\int_0^1 d\tau\, \frac{\Phi_V(\tau)}{\tau(1-\tau)} \,.
\label{sub-amp} \ee The meson's decay constant and distribution
amplitude are denoted by $f_V$ and $\Phi_V$, respectively. A model
for the gluonic GPD, $H^g$, is constructed by starting from a
double distribution \cite{rad99a} and writing it as a product of
the ordinary gluon distribution $x g(x)$, a factor $f(t)$ that
models the $t$-dependence, and an additional $x,y$ dependence

\be f^g(x,y,t) = 6 \frac{y(1-x-y)}{(1-x)^3}\, x g(x) \, f(t)\,.
\label{f1} \ee Using results given for instance in \cite{rad99a},
$H^g$ can be straightforwardly worked out from (\ref{f1}). It is
obviously related to the ordinary parton distribution which, for
small $x$, can be approximated by
 \be
x g(x,Q^2)\simeq 1.94 x^{-0.17-0.05 \ln(Q^2/Q_0^2)}\,,
\ee
where $Q_0^2=4\gev^2$.


Let us now turn to the modifiation of the leading-twist amplitude
in a way analogously to the treatment of the electromagnetic meson
form factors \cite{li92,jak93}. The basic idea is to retain the
transverse momenta of quark and antiquark forming the meson (
defined with respect to the meson's momentum). According to
phenomenological experience these transverse momenta are large
since the valence Fock state of the meson is rather compact. On
the other hand, the transverse momenta of the gluons (defined with
respect to the proton momenta) emitted and reabsorbed by the
protons, are much smaller because the full proton with its large
radius is to be considered. In contrast to \cite{vander} we
neglect the latter transverse momenta. Our approach is is similar
in spirit to the modification of the two-gluon exchange model
proposed in \cite{fra95}.

Taking into account transverse quark momenta, we have to replace
the distribution amplitude $\Phi_V$  by a soft light-cone wave
function for which we adopt a Gaussian parameterisation
\cite{jak93} \be
          \Psi_V(\tau,\vk)\,=\, 8\pi^2\sqrt{2N_c}\, f_V\, a^2_V
       \, \exp{\left[-a^2_V\, \frac{\vk^{\,2}}{\tau(1-\tau)}\right]}\,.
\label{wave-l} \ee

Transverse momentum integration of (\ref{wave-l}) leads  to the
associated distribution amplitude $\Phi_V^{AS}=6\tau\bar{\tau}$,
the asymptotic form of a meson distribution amplitude. For the
decay constants $f_V$ we take the values $0.216 {\rm GeV}$ and
$0.237{\rm GeV} $ for $\rho$ and $\phi$ mesons, respectively. For
the transverse size parameter $a_\rho$ we use a value of $0.6
\,\gev^{-1}$ which leads to a r.m.s. transverse momentum of
$0.6\gev$ and a valence Fock state probability of about 1/4, the
same value as for the pion \cite{jak93}. In the case of $\phi$-
meson production we take $a_\phi=0.45 \,\gev^{-1}$.

In the quark propagators of the subprocess amplitude we have to
consider the transverse momenta too. A typical modification of a
quark propagator is then

\be \frac{1}{\tau Q^2} \to \frac{1}{\tau Q^2+\vk^2}\,.
\label{m-prop} \ee
 For
consistency the inclusion of transverse degrees of freedom is to
be accompanied by Sudakov suppressions. Since the Sudakov factor
$S$ exponentiates in the transverse separation space it is
convenient to work in $b$ space. Fourier transforming the wave
function and the subprocess amplitude and collecting all terms, we
obtain for the gluonic contribution to the helicity amplitude of
vector meson photoproduction at small $\xbj$
\ba {\cal
M}^{V(g)}_{0+,0+} &=& {e}\, {\cal C}_V\,
                \sqrt{1-\xi^2}\,(1+\xi)\,
                   \int d \bar{x} d\tau\, \tau (1-\tau)\, H^g(\xb,\xi,t) \nn\\[1em]
      &\times& \int d^{\,2} \vbs\, \hat{\Psi} (\tau, -\vbs)\,
                           \hat{{\cal H}}_g (\xb,\tau,Q,\vbs)\, \exp{[-S(\tau,b,Q)]}\,.
\ea
Here, $\hat{{\cal H}}_g$ and $\hat{\Psi}$ are the plural
Fourier transforms of the subprocess amplitude and wave function,
respectively. The Sudakov factor has been calculated by Botts and
Sterman \cite{botts} to next-to-leading-log approximation using
resummation techniques; its explicit form can  be found for
instance in \cite{dahm}. The Fourier transform of   the subprocess
amplitude leads to Bessel functions  like $K_0(\sqrt{\tau}Qb)$.
\\[.4cm]

\begin{minipage}{5.4cm}
\phantom{aa} \hspace{-.4cm}\epsfxsize=5.5cm\epsfysize=5.2cm
\centerline{\epsfbox{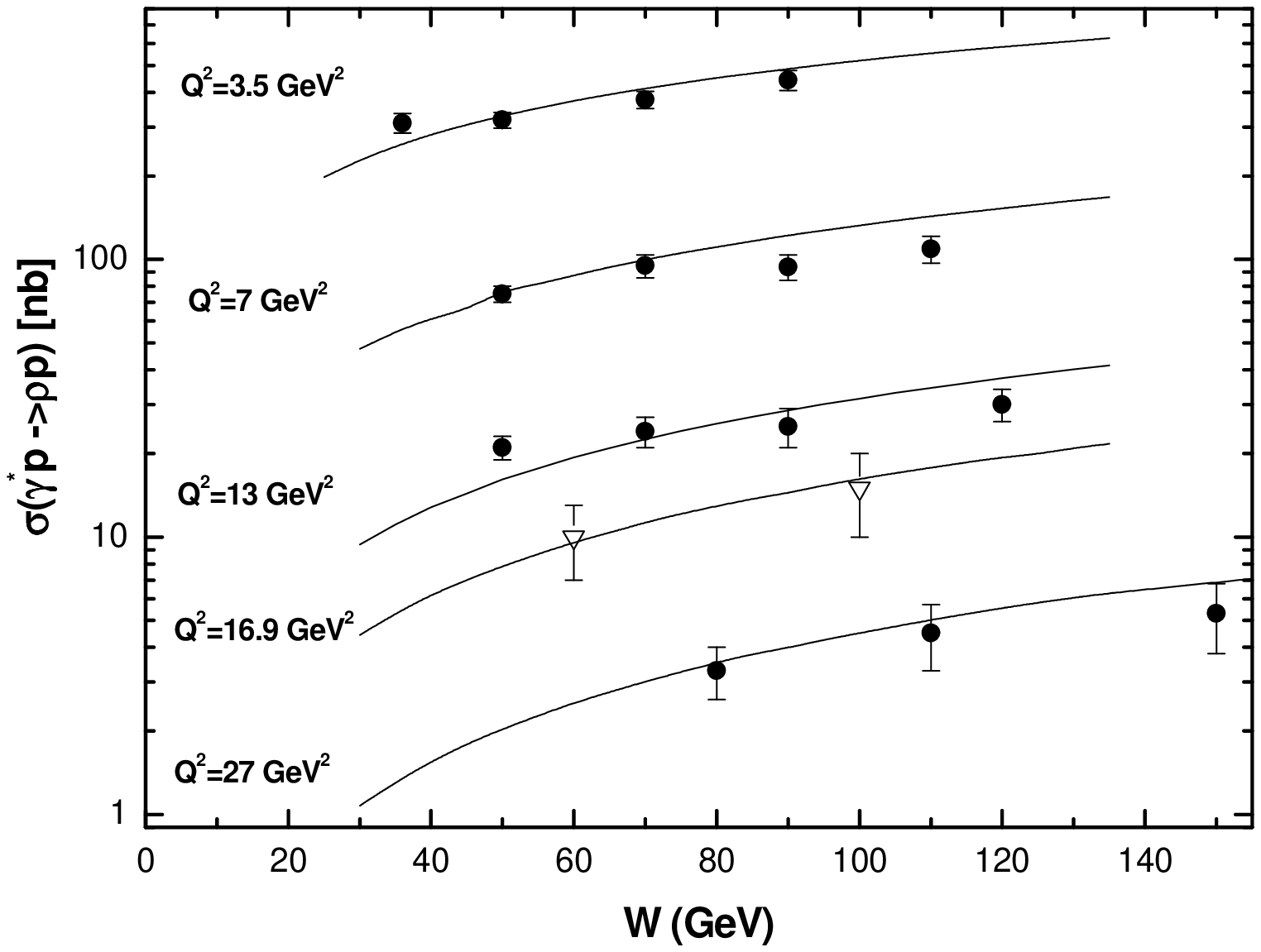}}
\end{minipage}
\begin{minipage}{0.1cm}
\phantom{aa}
\end{minipage}
\begin{minipage}{5.4cm}
\phantom{aa} \hspace{-.3cm} \epsfxsize=5.6cm \epsfysize=5.1cm
\centerline{\epsfbox{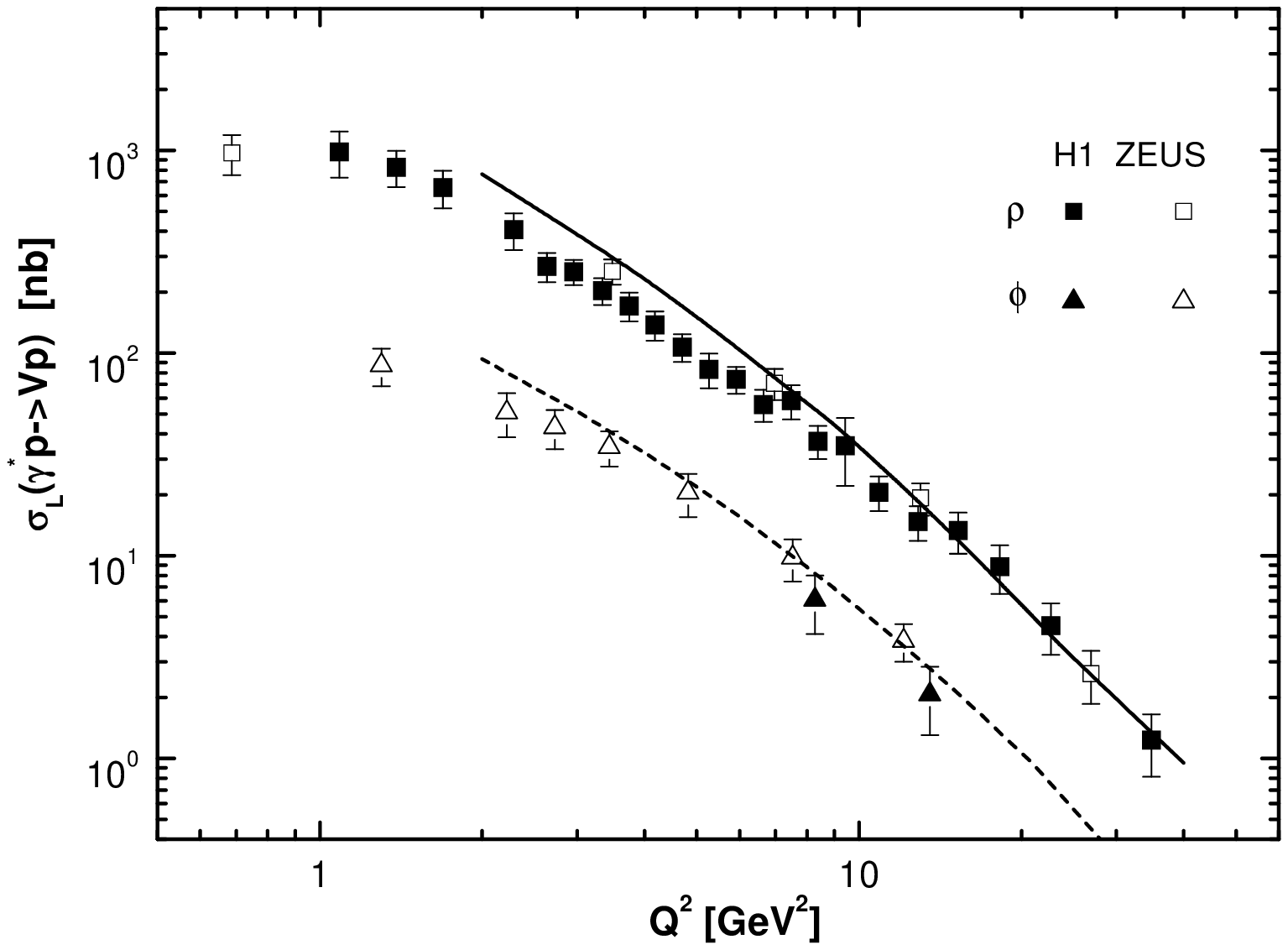}}
\end{minipage}
\\[3mm]
\begin{minipage}{5.6cm}
Fig. 1.~ {The cross section for $\gamma^*\, p\to \rho^0\, p$ vs.
$W$  for fixed values of $Q^2$.  Data are taken from
\cite{zeusdata}; ZEUS(95) open triangles, ZEUS(98) full circles.}
\end{minipage}
\begin{minipage}{.1cm}
\phantom{aaa}
\end{minipage}
\begin{minipage}{5.6cm}
Fig. 2.~ {The longitudinal cross sections  at $W=75 {\rm GeV}$
extracted in \cite{zeus}.
   The solid (dashed) line represent our results for
   $\rho$ ($\phi$) production. }
\end{minipage}
\\[5mm]


Our results for the $\gamma^* p \to \rho p$ cross section are shown on
Fig.1. The full cross section has been obtained from the
longitudinal one by adding to it the transverse contribution through
\be
\sigma(\gamma^* p \to \rho p)=\sigma_L(\gamma^* p \to \rho p)\,
(\epsilon+\frac{1}{R})\,.
\ee
A value of 3 for the ratio of longitudinal over transversal
cross sections and $\epsilon \simeq 1$ has been used by us.
As can be seen
from Fig.\ 1 good agreement of our approach with experiment
\cite{zeusdata} is obtained for a large range of $W$ and $Q^2$.

In Fig.\ 2 we show the longitudinal cross  sections for $\rho$ and
$\phi$ production  extracted from $\gamma^* p \to V p$ in
\cite{zeus} (references to the H1   and ZEUS data can be found
therein).  Again good agreement with experiment is to be observed
in the case of $\rho$ as well as in that of $\phi$ meson. We note
that our results for $\rho$ and $\phi$ production approximately
scale as the squared flavor factors, $({\cal C}_\phi/{\cal
C}_\rho)^2$. This comes about as a consequence of the quark
charges in the mesons and the similarity of the wave functions for
the light vector mesons.

We conclude -- the GPD approach to vector meson leptoproduction at
small $\xbj$,  modified by the inclusion of tranverse momenta in
the meson wave function and Sudakov suppressions, leads to
reasonable agreement between theory and experiment for   $\rho$
and $\phi$ production  cross sections. It is worth mentioning that
our approach can also be applied to $\gamma^*_T\to V_L$ and
$\gamma^*_T\to V_T$ transitions. The infrared divergencies
occuring in these transition amplitudes which signal the breakdown
of factorization \cite{man} are regulated by by the quark
transverse momenta.

\end{document}